**Rotationally-Resolved Spectroscopic Characterization of near-Earth object (3200) Phaethon**

[1]Theodore Kareta, [1]Vishnu Reddy, [1]Carl Hergenrother, [1]Dante S. Lauretta, [2]Tomoko Arai, [3]Driss Takir, [4]Juan Sanchez, [5]Josef Hanuš

[1]Lunar and Planetary Laboratory, University of Arizona, 1629 E University Blvd, Tucson, AZ 85721-0092

[2]Planetary Exploration Research Center, Chiba Institute of Technology, 2-17-1, Tsudanuma, Narashino City, Chiba Prefecture 275-0016

[3]NASA Johnson Space Center, 2101 NASA Parkway, Houston, TX 77058

[4]PSI, 1700 East Fort Lowell, Suite 106 * Tucson, AZ 85719-2395

[5]Institute of Astronomy, Charles University, Prague, V Hole\v sovi\v ck\'ach 2, CZ-18000, Prague 8, Czech Republic

Editorial Correspondence to:

Theodore Kareta

Lunar and Planetary Laboratory, University of Arizona

1629 E University Blvd, Tucson, AZ 85721-0092

tkareta@lpl.arizona.edu

Abstract:

(3200) Phaethon is a compelling object as it has an asteroidal appearance and spectrum, produces a weak dust tail during perihelion at just 0.14 AU, and is the parent body of the Geminid Meteor Shower. A better understanding of the physical properties of Phaethon is needed to understand the nature of its current and previous activity, relationship to potential source populations, and to plan for the upcoming flyby of the DESTINY+ spacecraft of Phaethon in the 2020s. We performed rotationally-resolved spectroscopy of Phaethon at visible and near-infrared wavelengths (0.4-2.5 microns) in 2007 and 2017, respectively, to better understand its surface properties. The visible and near-infrared observations both spanned nearly a full rotation or more and were under similar observing geometries, covering the whole surface with the exception of the north pole. The visible wavelengths show blue slopes with only minor slope variations and no absorption features. The NIR data is minimally varying and concave upwards, from very blue to blue-neutral with increasing wavelength. We fit the short-wavelength tail of Phaethon's thermal emission and retrieve an average visible albedo of $p_v$ = 0.08 +/- 0.01, which is lower than previous measurements but plausible in light of the recent larger radar-measured diameter of Phaethon. We retrieve an average infrared beaming parameter of Phaethon of $\eta$ = 1.70 +/- 0.05, which is similar to previous results. We discuss the implications of Phaethon's visible and near-infrared spectrum as well as the lower albedo on its origin, source population, and evolutionary history.

1. Introduction

1.1 Known Properties of (3200) Phaethon

(3200) Phaethon is an Apollo-type Near Earth Asteroid (NEA) that was noted quickly after discovery to be associated dynamically with the Geminid meteoroid stream (Whipple, 1983). With a semimajor axis of 1.271 AU, an eccentricity of 0.889, its perihelion distance (q = 0.14 AU) is one of the smallest known, setting aside the sungrazing comets seen in coronagraphic images (e.g., Marsden 2005). After decades of searching, possible activity was first observed on Phaethon by the sun-observing STEREO spacecraft (Jewitt and Li, 2010, Li and Jewitt, 2013). The comet-like orbit, when combined with its asteroidal appearance and somewhat enigmatic 'out-dusting' at perihelion, has led to it being called a 'rock comet' (Jewitt and Li, 2010). Before the most recent apparition, estimates from lightcurve inversion and thermal modeling have estimated the diameter of the object to be ~5.1+/- 0.2 km, and the rotational period to be ~3.6 hours (Hanus et al., 2016, and references therein.)

Phaethon is also associated with two other NEAs: (155140) 2005 UD and (225416) 1999 YC. Taxonomically, Phaethon and 2005 UD (e.g., Kinoshita et al., 2007) are both classified as B-types (in the DeMeo et al., 2009 taxonomy), while 1999 YC showed redder colors more indicative of a C-type (Kasuga and Jewitt, 2008). B-types, like other primitive asteroid types, are often associated with carbonaceous chondrite meteorites based on their near-infrared (NIR) reflectance spectra (see Clark et al., 2010, and references therein) – though exact curve matching of meteorite to asteroid spectra can be more challenging than with C-types. In particular, B-types are often best described as heated or "anomalous" carbonaceous chondrites. A relevant recent example is that of Clark et al. (2011)'s study of the target of the OSIRIS-REx mission, the B-type asteroid Bennu (101955). They found that the spectrum of Bennu was best fit by a sample of the CI Chondrite Ivuna, which had been heated to over 1000 K (Clark et al., 2011), which is not very different from Phaethon's current thermal environment near perihelion.

Phaethon is also the target of the Japanese Aerospace Exploration Agency's (JAXA) DESTINY+ mission that plans a high-speed fly-by following its launch in 2022 (Arai et al., 2018). The mission would study the dust environment around Phaethon and 2005 UD. As a result, ground-and space-based characterization of Phaethon ahead of the flyby is critical for planning the high-resolution imaging sequences and understanding the hazard environment around the object.

1.2 – Proposed Origin Scenarios and Other Recent Relevant Work

Phaethon's high eccentricity and associated meteoroid stream initially implied a cometary origin (e.g., Whipple 1983), but recent work mainly points towards an origin in the Main Belt. In their study of NEO orbits and source regions, Bottke et al. (2002) note that Phaethon has an 80% chance of coming from the inner belt and a 20% change of originating in the middle parts of the main belt. However, their model cannot reproduce the orbits of objects like the Jupiter Family Comet 2P/Encke, which has a low perihelion distance (~0.33 AU) and a semimajor axis in the main belt. (Thus, the possibility of it being a dead comet perturbed by the terrestrial planets is possible but unlikely.) In their study of the orbit of Comet Encke, Levison et al. (2006) show that a plausible pathway exists for Jupiter Family Comets (JFCs) to become decoupled from Jupiter, spend time as an inactive comet in the main belt, and eventually reactivate when its perihelion distance lowers sufficiently. However, they note that objects with Tisserand Parameters above 3 (Phaethon has 4.51) are still far more likely to be from the main belt than JFCs, which have become decoupled from Jupiter (Levison et al., 2006).

Licandro et al. (2007) and Clark et al. (2010) both obtained VNIR observations (0.4-2.45 microns) of Phaethon and compared their data to meteorites, laboratory samples of minerals and artificial substances, as well as spectral mixture models. Licandro et al. (2007) find their spectrum is best fit by either heated carbonaceous chondrites (a RELAB heated sample of Ivuna or the anomalous CI/CM chondrite Y86720) or a modeled synthetic mixture of lampblack and a hydrated silicate (Montmorillonite or Antigorite). Clark et al. (2010) best fit their spectrum with a CK4 chondrite or a combination of a hydrated silicate (chlorite) and lampblack. de Leon et al. (2012) also found that a CK chondrite is the best fit for Phaethon. More recent work by Takir et al. (2018) has shown that Phaethon lacks a 3-micron feature suggesting a surface regolith that is devoid of any hydrated silicates.

Vernazza et al. (2015) proposed that the best analog for C-complex asteroids was not meteorites, but interplanetary dust particles (IDPs) instead. A visible and near-infrared reflectance spectrum of Phaethon was fit quite well using IDPs as the basis of a radiative transfer model. Broadly speaking, the asteroid/comet classification issue for Phaethon becomes less important within this framework, as they propose (Vernazza et al., 2015) that both C-complex asteroids and comets should both be made out of anhydrous IDPs, which agrees Phaethon's lack of hydrated surface minerals (Takir et al., 2018). We also note that the ample ~1-micron sized dust particles needed to explain the perihelion brightening of Phaethon (Jewitt and Li, 2010) is similar in size to some IDPs.

de Leon et al. (2010) claim that the most likely parent body for asteroid Phaethon is the large main-belt B-type asteroid (2) Pallas. This association is based on the spectral similarity between Pallas and Phaethon in the NIR and between Pallas Collisional Family (PCF) members and Phaethon in the visible wavelengths. This pairing also helps to explain the high visible albedos of Phaethon ($p_v$ = ~0.12, Hanus et al. 2016) and Pallas ($p_v$ = ~0.16 for Pallas, Tedesco et al., 2002), compared to typical B-types. Additionally, dynamical analysis shows a pathway for Phaethon to evolve from a Pallas-like orbit to highly eccentric and moderately inclined orbits in near-Earth space. Recently, Todorovic (2018) did more rigorous dynamical work showing that the pathway from PCF-like orbits to Phaethon-like orbits was even more efficient than previously thought.

Several recent studies have analyzed the ensemble properties of the PCF and B-types in the main asteroid belt (e.g., Ali-Lagoa et al., 2013, Ali-Lagoa et al., 2016). The PCF has, on average, a higher visible albedo (0.14 +/- 0.05) than other B-types of similar size (0.07 +/- 0.02) (Ali-Lagoa et al., 2013, 2016). The same work, when comparing asteroids of similar sizes, found that PCF members had lower thermal-beaming parameters than other B-types, indicating a difference in surface properties assuming that asteroids of the same size have similar regolith thickness. Phaethon's previously reported albedos of 0.10-0.13 (e.g., Tedesco et al., 2002, Hanus et al., 2016) is thus slightly higher for an average main belt B-type and somewhat low for an average PCF member of similar size. Harris (1998) found a best-fit beaming parameter of 1.6 for Phaethon, which is higher than any PCF member (maximum $\eta$: 1.5, average $\eta$: ~1) whose beaming parameter fit was considered in Ali-Lagoa et al. (2016), with the discrepancy becoming larger when comparing like-sized objects. Furthermore, in a broad spectroscopic study of B-types including Themis and Pallas Collisional Family members, de Leon et al. (2012) showed that B-type spectra largely fall on a continuum of red-to-blue NIR slopes, where Phaethon is the bluest end-member. Thus, the difference between the Pallas and Themis family spectra becomes less distinct. While Phaethon's spectral similarity to Pallas and members of the PCF is clear, it is unclear that this correlation is as diagnostic as previously perceived (de Leon et al. 2010) given the observed continuum of NIR slopes. Furthermore, the recent revelation that Phaethon's surface is spectroscopically featureless at 3.0 microns (Takir et al., 2018), unlike Pallas, is another discrepancy between the two objects that needs to be better understood.

Much recent work focuses on understanding the current activity of Phaethon. The original reporting on the activity on the object (Jewitt and Li, 2010, Li and Jewitt, 2013) was a ~2 magnitude brightening as it approached perihelion in the STEREO images, consistent with forward-scattering of solar light by ~1-micron size dust particles. Jewitt et al. (2013) report that the mass loss through a single perihelion passage (~3 x $10^5$ kg) is dwarfed by the mass of the Geminid stream (~$10^{12}$-$10^{12}$ kg, Jenniskens 1994), indicating that the current activity of Phaethon must be weaker (and perhaps different). Recent literature has started to coalesce around the idea that thermal decomposition of minerals (namely phyllosilicates) and thermal fatigue (Jewitt and Li, 2010) produced the dust in Phaethon's perihelion dust tail. The dust can then be removed from the surface by some combination of radiation pressure, electrostatic levitation, and rotational effects.

1.3 – The December 2017 Close Approach and Observational Campaign of (3200) Phaethon

We observed Phaethon as part of a global campaign during a close approach to the Earth, centered around December $16^{th}$, 2017, when the asteroid was 0.069 AU away. Radar observations of Phaethon were conducted at the Arecibo Observatory (Taylor et al., 2018), which revealed the body to be somewhat larger (D ~ 5.7 km) than previously estimated from thermal studies (e.g. 5.1 +/- 0.2 km from Hanus et al., 2016, 4.2 +/- 0.13 km from Usui et al., 2011). Radar observations are known to be highly reliable in reproducing asteroid sizes and shapes.

However, Wooden et al. (2018) have reported a diameter of D ~ 4.1 km derived from new speckle imaging, which is somewhat lower. Further detailed work will be needed to understand the (possibly growing) discrepancy between size estimates for Phaethon.

Furthermore, polarimetric measurements made during the 2017 apparition revealed Phaethon to have extremely strong linear polarization at high phase angles, either indicating a lower albedo than previously reported (Ito et al., 2018, Devogèle et al., 2018, Borisov et al., 2018) or a surface with rare polarimetric properties (e.g., a lack of small grains) or some combination of the two. In short, Phaethon deviates strongly from Umow's Law, which relates the geometric albedo of an object to the maximum level of linear polarization. The new radar-derived diameter implies the albedo must be lower than previously reported, but any other new constraints on the albedo would also thus allow constraints on physical properties of the surface that create such a robust polarimetric response. Additionally, while Phaethon's polarization response is very strong, it seems to be similar in behavior to other B-types, including Pallas, as opposed to F-types (Devogèle et al., 2018).

Our primary motivation for observing the asteroid was to provide the best understanding of this enigmatic object until DESTINY$^+$ visits it in the 2020s. In this paper, we present results of our rotationally resolved near-IR reflectance spectroscopy study of (3200) Phaethon from the December 2017 close approach. We also used archival rotationally resolved visible wavelength spectral data (0.4-0.74 microns) from November 2007 apparition to complement our NIR dataset. We discuss the implications of our observations on the surface properties and history of Phaethon.

2 – Observations and Data Reduction

[Table 1]

We conducted visible wavelength observations from 0.4 to 0.74 microns on November 5[th], 2007 (UTC) from the Tillinghast 1.5-meter telescope on Mount Hopkins, Arizona using the high-throughput FAST spectrograph (Fabricant et al., 1998), which provides a spectral resolution of R ~1360. As all observations of the asteroid were within 0.22 airmass of each other, we observed three nearby standard stars at a range of airmass values. The sub-observer location was ~117 degrees from the rotational pole derived by Hanus et al. (2016), so part of the northern hemisphere was not visible. After standard wavelength and flat-field calibration, we divided our asteroid spectra by the spectra of the standard star. The standard is the same for both our visible and NIR observations and provides a consistent dataset.

We conducted infrared observations from 0.7 to 2.5 microns on December 12[th] (UTC), 2017 with the SpeX instrument (Rayner et al., 2003) at the NASA Infrared Telescope Facility (IRTF) on Mauna Kea, Hawaii (See table of observing circumstances). The sub-observer location was ~122 degrees from the rotational pole, similar to our visible observations. Observations of the asteroid were 'bookended' by observations of a G-type standard star close to the asteroid on the sky at similar airmass. Later, a primary solar-analog star was observed to account for differences between the standard star spectrum and the solar analog. We collected data with the slit oriented to the parallactic angle to prevent wavelength dependent differential refraction. The data were reduced with Spextool (Cushing et al., 2004), a set of interactive IDL scripts and GUIs written for users of the instrument. A detailed description of our data reduction protocol is presented in Reddy (2009).

3 – Results

[Figure 1]

The combined average visible and near-IR (0.4-2.55 microns) spectrum of Phaethon is shown in Figure 1 normalized at 0.55 microns. The following trends are evident. The visible (0.37-0.7 micron) spectrum is featureless but has a nearly uniform blue spectral slope. There is a further decrease (bluer) in slope around ~0.7-0.75 microns, which is near the short-wavelength end of the NIR data and the long-wavelength end of the VIS data. The spectrum is generally concave upwards through the NIR after ~1.2-1.3 microns, with an incredibly blue slope at 1 micron slowly increasing to a gentle blue slope by ~1.8 microns. The thermal tail (exponential-like upturn) is visible beyond ~2.0 microns. Previous observations of Phaethon taken at further heliocentric distances do not show this thermal tail.

[Figure 2, 3]

The rotationally resolved visible wavelength spectra are compared in Figure 2, and the best-fit slopes for our visible observations are shown in Figure 3. The visible and infrared datasets were captured to record the spectral features of the asteroid across a full rotation. As the radar-derived shape information has only been published in a conference abstract (Taylor et al., 2018), we use the rotational information presented in Hanus et al. (2016) to calculate the rotational phases of our observations. The visible observations start at 0.499 rotational phase and end slightly less than one full rotation later at 1.432, for a total of 0.937 rotations. Our NIR observations start at 0.547 rotational phase and end more than one full rotation later at 1.639, for a total of approximately 1.09 rotations.

In the absence of diagnostic absorption features in the reflectance spectrum, spectral slope could be used as a non-diagnostic parameter when searching for rotational variations on an asteroid. Our slope fitting was primarily accomplished through the SciPy 'curve_fit' routine (Jones et al., 2001), a chi-squared minimization algorithm, utilizing a linear fit to model the data between two input bounds in wavelength. For our visible wavelength data, we fit our entire spectrum with one line, as well as our data shortwards and longwards of 0.55 μm to search for any possible subtle changes in slope despite the highly linear appearance of the spectrum (Figure 2). For our NIR data, we fit 0.75-1.1, 1.1-1.45, and 1.45-1.8 μm for similar reasons. The data longwards of that are dominated by the 1.9-μm telluric band followed by the thermal tail, and we treat them separately. Generally speaking, there are slight variations in the overall slope throughout the visible wavelengths, with larger variations in the smaller wavelength bins. The spectrum of Phaethon appears to get slightly less blue near ~0.4-0.6 phase. We see no evidence for a decrease in reflectance at shorter wavelengths, as well as no evidence for transient or persistent absorption features. A single red-sloped spectrum near ~0 phase dominates that redder slope region, though we also note that this particular spectrum is much noisier than our other data and thus suggest it is likely not representative of a real change in slope. In general, the data are consistently blue and only vary in intensity with the possible exception of a noisy spectral bin near phase ~0.0 and another spectral bin near phase ~0.4 for the shortest wavelengths considered.

[Figure 4, 5]

We compare the rotationally resolved NIR observations in Figures 4 and 5. Similarly, the rotational variation in spectral slope is significant and much larger in magnitude at shorter wavelengths. The 'curved' nature of the NIR spectrum of Phaethon, where the shortest wavelengths are bluest, is evident in both the spectral comparison and slope comparison plots. We note here that around ~0.5 phase there is a more significant variation in slopes which is contemporaneous with changes in humidity at the observing site, so it should be interpreted cautiously.

3.1 – Thermal Modeling

**[Figure 6, 7]**

Many near-Earth objects, especially those with low albedo, have appreciable thermal emission beyond 2.0 microns when observed at small heliocentric distances, which need to be corrected away to inspect the underlying reflectance spectra. This correction, usually concerning the application of a thermal model, also allows for estimation of the thermal properties of the asteroid. A thermal upturn occurs in all of our NIR data (see Figure 4), which is unsurprising considering Phaethon is a low-albedo asteroid which we observed at ~1.08 AU. In Figure 6, we display one of our thermal models fitted to our data, and we show our best-fit thermal albedos as a function of the rotation phase in Figure 7. Due to increased scatter at long wavelengths during changing humidity conditions, we only fit our data, which were taken during stable observing conditions.

We utilized the Near-Earth Asteroid Thermal Model ("NEATM," Harris 1998), which is, in turn, a modification of the Standard Thermal Model ("STM," Lebofsky et al., 1986). For our purposes, the model has two open parameters: the albedo in the visible band ("$p_v$") and the infrared beaming parameter ("$\eta$"), both of which adjust the amount of the observed thermal emission. We note here that we were unable to fit the data satisfactorily with beaming parameters lower than 1.5, which is incompatible with the STM. We largely follow the implementation of the NEATM described in Rivkin et al., 2005 and Reddy et al., 2009 whereby we fit the "thermal excess" Y as a function of the thermal emission T and the reflected light R:

$$Y(\lambda) = \frac{T(\lambda) + R(\lambda)}{R(\lambda)} - 1$$

This ratio allows us to essentially fit the 'shape' of the thermal emission, instead of absolute fluxes. The lack of flux-calibrated data reduces our ability to constrain diameter and albedo independently, and as a result, the thermal excess was originally constructed to be independent of diameter, as it is a part of both the reflected and emitted light.

The reflected light is modeled as:

$$R_{model}(\lambda) = D^2 F_{Sun}(\lambda) R^2 \Delta^2 \phi(G) p(\lambda)$$

$$p(\lambda) = R_{model-uncalibrated}(\lambda)\, p_v\, f_{kv}$$

where D is in kilometers, R in astronomical units (AU), Δ in kilometers. $\phi$ and p are unitless.

We have followed the convention that we model the underlying reflectance of the asteroid ("$R_{model-uncalibrated}$") as a straight line fit to a shorter-wavelength segment of the reflectance spectrum (e.g., Rivkin et al., 2005, Reddy et al., 2009). However, this approach can allow misinterpretation of the thermal excess if the underlying reflectance is not changing linearly or the range over which the continuum is fit is unrepresentative of the underlying reflectance. Phaethon has been noted previously (e.g., Licandro et al., 2007) to have a spectrum that is slightly curved, with wavelengths further into the infrared generally displaying a less blue slope (though never genuinely becoming red). The non-linear nature of Phaethon's spectrum in the NIR makes estimating the underlying reflectance continua more challenging than with other objects.

Our model reflectance continua were chosen to minimize the chi-squared value of the final fit while also correctly predicting thermal excess values close to zero in the non-thermal regime. The spectral slopes inferred for each continuum fit ranged between -0.04 and +0.01 per micron, which is consistent with slopes measured from previous observations of Phaethon made at larger heliocentric distances, such as in the SMASS database. We were also able to obtain similar, but noisier fits using SMASS spectra as our underlying reflectance assumption, but have not included those here. While the range used to fit the underlying reflectance continuum varied to allow for quality fits, we only used data between 1.92 and 2.5 microns in fitting the thermal parameters.

$p_v$ is the visible wavelength (0.55 micron) albedo, and $f_{kv}$ is the ratio of $p_k$, the k-band albedo, to $p_v$, which we estimate using a combination of our combined VIS+NIR dataset and our best-fit underlying reflectance (see above) in the thermal region. Among our thermal fits that converged, $f_{kv}$ was between 0.71-0.72. $\phi$, the phase integral, we calculate assuming the G value (Bowell et al., 1989) of 0.06 measured by Ansdell et al. (2014), a common value for darker objects. As an aside, Hanus et al. (2016) derived that G = 0.15 +/- 0.03, which we discuss further in our Discussion further. As both our modeled thermal emission and reflectance are a function of the diameter squared, it disappears from our modeled thermal excess coefficient. The NEATM model also assumes that thermal emission (and reflected light, in the Reddy et al., 2009 implementation) only comes from the illuminated fraction of the object. We observed Phaethon in the NIR at a phase angle of ~22 degrees, which results in a relatively small correction.

Among the nine thermal fits considered the average albedo is 0.08 +/- 0.01, with individual fit values ranging from 0.052 to 0.11. The average beaming parameter is 1.70 +/- 0.05, with individual values ranging from 1.52 to 1.97. Generally, the noise for each fit is high enough to make discerning trends throughout the rotation of Phaethon challenging, especially when combined with the lack of fits during the period of poor observing conditions. As mentioned previously, beaming parameters close to 1.0 (or lower, in case of the original STM) produced extremely poor fits with implausibly high albedo estimates ($p_v > 0.25$).

4.0 – Discussion

4.1 Albedo and Thermal Properties:

The radar-derived diameter of 5.7 km (Taylor et al., 2018) is the largest estimate for Phaethon yet, and thus requires a revision of other known properties of Phaethon. In particular, the previously derived diameters were primarily radiometric, where the visible albedo (as a proxy for the geometric albedo) and the diameter are related as (Fowler and Chillemi 1992):

$$p_v = (1329 / D[km])^2 * (10^{-H/5})^2$$

As the new diameter is larger than assumed previously, one possibility is that the visible albedo, therefore, is lower. Our average best-fit albedo (0.08 +/- 0.01) is fully consistent with what one would estimate from the diameter-absolute magnitude equation using the radar-derived diameter, and the dimmer of the two reported absolute magnitudes (14.6, from JPL HORIZONS), which results in an estimate of $p_v$ = 0.078. Using H=14.3 (e.g., Hanus et al., 2016), we retrieve $p_v$ = 0.104, which is 2 sigma higher than our reported average albedo. Generally speaking, we prefer an absolute magnitude value on the dimmer end of the published values.

Unlike H, the slope parameter G is an input to our model and thus has a weak effect on our final output albedo estimate. Hanus et al. (2016) estimate a higher value of G = 0.15 +/- 0.03, which if utilized in our model would retrieve an even lower average albedo of $p_v$ = 0.068 +/- 0.009. However, the currently publically available Minor Planet Center photometric dataset generally supports a value of G closer to Ansdell et al. (2014)'s G = 0.06 rather than Hanus et al. (2016)'s higher value (Hergenrother, personal communication).

In fitting our data, we noticed a definite trend that only specific small ranges of modeled reflectance (almost always slightly blue, which is consistent with many previous observations) allowed for high-quality fits to the data. In other words, the underlying reflectance could change slightly, and a high-quality fit with very similar thermal fit parameters could be retrieved. However, this possible variation is dwarfed by the spread in retrieved albedos for our nine fits. Even considering all these caveats, we consider our results to be a strong indication that the actual albedo of Phaethon is lower than previously reported.

Our average best-fit albedo is significantly lower than the original IRAS-derived albedo (0.11 +/- 0.01, Tedesco et al., 2002) and the more recent Spitzer-derived albedo (0.122 +/- 0.008, Hanus et al., 2016). The IRAS result utilized the Standard Thermal Model, which has been shown to not work very well on smaller or near-Earth objects. Furthermore, both the IRAS and Hanus et al. (2016) results would be affected significantly if the reported absolute magnitude were too bright, which is typical for dark NEOs or asteroids with incomplete phase angle coverage, both of which are true for Phaethon. Our data is thus an independent suggestion that the real absolute magnitude of Phaethon is the dimmer of the two reported magnitudes (14.6, JPL Horizons), or perhaps even slightly dimmer.

The albedo of a featureless asteroid is useful in identifying its taxonomic type. B-type asteroids outside of the Pallas Collisional Family have albedos in the range of 0.07 +/- 0.02 (Ali-Lagoa et al., 2016), our average best-fit of the albedo of Phaethon (0.08 +/- 0.01) clearly fits within this range. The large albedo discrepancy between Phaethon and Pallas (0.16 +/- 0.01, Tedesco et al. 2002) complicates the hypothesis that Phaethon is a collisional fragment from the PCF.

The recent polarimetric observations of Phaethon (Ito et al., 2018, Devogèle et al., 2018, Borisov et al., 2018) reveal it to deviate strongly from the relationship between albedo and the maximum degree of linear polarization ("Umow's Law") if we use one of the previously published albedos. This deviation can be explained by a lower albedo, a surface with rare polarimetric properties, or some combination of both. Our data generally support the first hypothesis, namely that Phaethon's albedo is lower (0.08 +/- 0.01) than previously reported value ($p_v$ = 0.122 +/- 0.008 without correction for the new diameter, $p_v$ = 0.10 +/- 0.01 with correction) from Hanus et al. (2016).

Our best thermal fits to the Phaethon data result in an average beaming parameter of $\eta$ = 1.70 +/- 0.05, which could be indicative of a surface not in radiative equilibrium – perhaps due to rotational state, partial regolith coverage, or non-negligible thermal inertia (see below). Harris (1998) applied NEATM to Phaethon and noted that it had an above-average best-fitting value for beaming parameter ($\eta$ = 1.6), which is very similar to ours. They suspected that Phaethon might have significant thermal emission from the night side or unusual thermal properties. Hanus et al. (2016) inferred the thermal inertia of Phaethon (Γ = 600 +/- 200 in SI units) from a detailed thermophysical model and found it slightly higher than average for the near-Earth asteroids, possibly implying a coarser regolith with a rougher surface.

4.2 Limitations of Our Modeling

Our application of the Near-Earth Asteroid Thermal model has several key limitations, which are worth mentioning in brief. First, the beaming parameter ("$\eta$") and visible-band albedo ("$p_v$") are known to be highly covariant. There could be a counterpart to our low albedo ($p_v$ = 0.08 +/- 0.01) and large beaming parameter ($\eta$ = 1.70 +/- 0.05) with a higher albedo and lower beaming parameter, though we were not able to retrieve a fit like this and did not see it in any of our chi-squared maps. The model is also fundamentally limited by its non-physical assumption that the thermal inertia of the asteroid is zero. Our NIR observations were taken at a relatively low phase angle (~22 degrees), so this issue is perhaps less important to our dataset as compared to other NEO observations, but should still be taken seriously. The model also implicitly assumes a spherical asteroid (at least in projection). Lastly, as our data is only of the short-wavelength tail of the thermal blackbody curve, our estimate of the relative contribution of the reflected light can have large effects on the resulting thermal fit if estimated incorrectly. As mentioned above, we were only able to retrieve plausible thermal fits for a relatively small range of reflectance slopes, so we find it unlikely that we chose incorrectly given the lack of other options. In general, a more detailed and physical thermal model will be able to determine the thermal properties of Phaethon to a much greater precision, which the new radar shape model of Phaethon (Taylor et al., 2018) should be very useful for. However, such a detailed thermal model is beyond the scope of this paper and the amount of thermal observations we have obtained for it.

4.3 Spectral Variations

Phaethon was initially classified as an F-type in the Tholen (1984) taxonomy, as it originally showed no evident UV absorption and more recently has been merged into the modern B-type (Demeo et al., 2009). Our visible data shows Phaethon to have a linear, uncurved spectrum with a consistently negative (blue) slope and no UV absorption. We see some evidence for a slight reddening near ~0.6 rotational phase, but no area which is 'red' or even neutral in slope. In other words, our data would classify Phaethon as a classical F-type and a modern B-type over its entire surface.

Clark et al. (2011) note that UV absorption on fine-grained material from Murchison samples from the RELAB database shows variations with phase angle. Our visible wavelength data is blue-sloped at a phase angle of ~41 degrees, which is inconsistent with Phaethon's surface regolith having optical properties similar to the Murchison matrix material, as the sample shown in that work has a red slope throughout the visible and is only blue at small phase angles.

Licandro et al. (2007) note that ground-based spectral observations of Phaethon have detected a UV absorption feature shortward of 0.55 μm with changing band depth. They hypothesize that the band-depth change might be due to a heterogeneous surface with grain size variation. In order for our dataset to be consistent with the previously reported observations, our visible observations must have been conducted nearly pole-on, such that the portion of the surface seen did not change significantly throughout the observations – but this is not the case.

Based on the best available shape and rotation information from Hanus et al. (2016), our observations spanned almost the entire surface of the asteroid. The large reported near-UV absorption variation thus cannot be explained by spatial variation – unless the northern rotational pole of Phaethon is significantly different from the rest of the asteroid spectroscopically. Interestingly, Borisov et al. (2018) also recently inferred that the northern pole may have different polarimetric properties. Furthermore, Taylor et al. (2018) note a radar-dark spot near one of the rotational poles as well. Further work is merited to understand how these lines of evidence are or are not related, and whether or not the north pole has different spectroscopic properties at blue wavelengths.

However, the Hanus et al. (2016) shape model does not suggest that any side of the asteroid is preferentially heated over another at successive perihelia. Any processes that are acting today to change the surface – be it thermal cycling of the rocks, equilibration of minerals, dust ejection by radiation pressure sweeping or another process – would act over the whole surface over a sufficiently long timescale of several orbital periods. Furthermore, we could interpret the subtle slope variations in our NIR data as evidence for a grain-size related effect. However, these grain-size variations would also change thermal properties in a way that might be observable with a more accurate characterization of the thermal emission from the asteroid than was possible with our data, as described above.

As mentioned previously, our near-infrared observations only show minimal variations in slope over a full rotation. A close inspection of the data presented in Figures 4 and 5 together will show that much of the variation comes from variation in the 'curve' of the spectrum itself. Phaethon was classified as nearly spectroscopically unique in de Leon et al. (2012) due in part to the intensity of this curve. All other B-type asteroids considered in that study had less 'blue' (negative) NIR slopes over the wavelength ranges considered, with many having neutral, more linear slopes and some being even red sloped (positive) over the wavelengths considered (de Leon et al., 2012). de Leon et al. (2012) contextualizes their B-type census in terms of meteorite analogs, suggesting that this feature is perhaps related to thermal processing, with Phaethon being the extremely heated end-member. Another set of possible origins of the different C-complex asteroid behaviors in the 1.0-1.3-micron range is a variation in the existence of amorphous olivine (Vernazza et al. 2015, 2017, Marsset et al., 2016) or other materials, like sulfides.

4.4 Source Region

In order for Phaethon to be a member of the Pallas Collisional Family (PCF), we should be able to explain deviations in its albedo and thermal inertia properties from the other PCF members in terms of its subsequent evolution. As mentioned previously, Pallas and the PCF members have characteristic spectra (e.g., de Leon et al. 2010, and references therein), albedos, and thermal beaming parameters (e.g., Ali-Lagoa et al., 2016, and references therein) that can aid in distinguishing possible ejected family members from other B-types. The dynamical work (de Leon et al., 2010, Todorovic, 2018) suggests that there are almost certainly PFC members in the NEO population with Phaethon like orbits. While the spectra of Phaethon, Pallas, and the PCF are strikingly similar, our measured albedo ($p_v$ = 0.08+/- 0.01) is lower than any measured value for a PCF member or Pallas (average $p_v$ ~0.12-0.14 (Ali-Lagoa et al., 2016), 0.16 +/- 0.1 (Tedesco et al., 2002), respectively.). Similarly, our inferred value of the infrared beaming parameter ($\eta$ = 1.70 +/- 0.05) is also higher than any observed beaming parameter for a PCF member (maximum $\eta$ = 1.5, average ~ 1, Ali-Lagoa et al., 2016).

One can interpret these differences in albedo and thermal inertia between Phaethon and PCF/Pallas in two ways. The first is that PCF/Pallas may not be the source of Phaethon, even if NEAs derived from the PCF are likely to be in Phaethon-like orbits. The second is that its surface properties have been modified since its arrival in near-Earth space and that differences between Phaethon and Pallas can be explained along these lines. Previous compositional modeling suggested that Phaethon was covered in hydrated minerals (Licandro et al., 2007), but recent work has shown the surface to be dehydrated (Takir et al., 2018), indicating surface alteration where the regolith is devolatilized compared to its main belt family and Pallas itself, which is hydrated. However, when studying the heating of water-bearing minerals, Hiroi and Zolensky (1999) found that there were significant spectral changes – from a general reddening trend to a loss of absorption features at 0.7 and 3.0 microns, among others. These results are a strong function of which hydrated silicate was studied and the grain sizes utilized. In light of this result, one would expect the thermally altered surface of Phaethon to have different spectral signature compared to PCF/Pallas. Hiroi and Zolensky (1999) also observed albedo changes where it decreased to 500-600 Celsius and increased as heating continued. This phenomenon may explain some of the deviations of Phaethon's measured albedo from the mean PCF albedo. As an additional note, some CI and CM chondrites begin to show a blue slope when heated – notably the CI Ivuna (Hiroi and Zolensky, 1999). More detailed lab work to understand how Pallas/PCF-like materials respond to Phaethon-like heating will help to unravel this mystery.

The question of the origin of Phaethon is fundamentally a question about the nature of its activity now and in the past. There is evidence suggesting it to be more active in the past (e.g., Jewitt and Li, 2010) and any near-surface volatiles, should they have existed, are almost certainly gone by now. Our new average best-fit albedo for Phaethon ($p_v$ = 0.08+/- 0.01) is in line with traditional B-type values (0.07 +/- 0.02 on average, Ali-Lagoa et al., 2016) and slightly larger than for cometary nuclei (0.02-0.06 on average, Lamy et al., 2004). For comparison, Comet 67P/Churyumov-Gerasimenko's visible albedo is similar but slightly lower at $p_v$ = 0.065 +/- 0.002 (Fornasier et al., 2015). Phaethon is, regardless of its relation to Pallas, most likely a primitive Main Belt Asteroid which was much more active on its current orbit at some point in the past, probably coinciding with the creation of the Geminids. The nature of its previous activity, while probably closer to traditionally cometary, is still unclear and merits more theoretical work.

Phaethon also shares specific characteristics with Jupiter Family Comets (JFC) such as 249P/Linear. 249P is a near-earth JFC that Fernandez et al. (2017) have shown is both spectroscopically a B-type and dynamically linked to the Main Belt. Most relevant to Phaethon is that its activity is only observable within about three weeks around perihelion and is on an orbit stable for ~$10^4$ years in Near-Earth space. We propose that further study of objects like 249P is necessary to understand Phaethon as it might have been near the end of its activity. In particular, work studying the current volatile content – such as by studying any possible emission from gases – would help to quantify what kind of volatile content Phaethon once had (or might still have) and understand the nature of its activity while the Geminid Meteoroid Stream was produced. Ideally, with further work, we could identify bodies throughout the entire evolutionary process that Phaethon has gone – from a likely source population (Pallas or otherwise) of volatile-rich main belt asteroids to a near-Earth comet with diminishing activity due its low perihelion distance like 249P to a mostly inactive object like Phaethon.

Summary

(3200) Phaethon is an enigmatic near-Earth object that has primarily asteroidal properties but transient near-perihelion activity and an associated meteor shower. In this work, we present and describe rotationally-resolved spectroscopic observations of Phaethon between 0.4 and 2.5 microns taken in 2007 and 2017. The surface spectrum is blue sloped throughout the visible wavelengths with subtle slope variations, which are more prominent at shorter wavelengths. In the NIR, the shortest wavelengths are seen to vary more than the longer wavelengths and to generally be much bluer. We see no absorption features at the ~7% level at visible wavelengths and at the ~1% or lower in the near infrared, including no UV absorption feature that has seen intermittently by other observers, and comment on the implications of this for Phaethon's composition and for spatial variation across its surface. We also analyze thermal emission seen at our longest wavelengths and find the visible wavelength albedo of the object to be on average $p_v$ = 0.08 +/- 0.01, which is significantly lower than measured by previous workers but plausible in light of recent radar observations. The infrared beaming parameter is on average $\eta$ = 1.70 +/- 0.05, which is consistent with previous studies. We argue against (2) Pallas as a likely source body for Phaethon, as spectral changes would accompany the albedo differences, and we comment on laboratory studies that would be helpful in understanding the relationship between the two bodies. Lastly, we note that Jupiter Family Comet 249P/Linear has many features analogous to what Phaethon might have been like in the recent past, making it a highly appealing target for future studies.

DESTINY+'s direct observations of the surface of Phaethon will allow resolution of this ambiguity. A spacecraft-derived shape model and a precise determination of the geometric albedo will result in more detailed thermophysical models to understand the processing of materials on the surface and in the interior, which will facilitate a better understanding

Phaethon's thermal history and how that history has shaped its modern properties. A more precise determination of albedo will also allow a better quantification on the deviation of Phaethon's surface materials from Umow's Law, which will allow for a deeper understanding of the scattering and reflectance properties of the regolith on Phaethon.

Acknowledgments

This research work was supported by NASA Solar System Observations Grant NNX14AL06G (PI: Reddy) and Arizona Technology Research Initiative Fund (TRIF). We thank the IRTF TAC for awarding time to this project, and to the IRTF TOs and MKSS staff for their support. Part of the data utilized in this publication were obtained and made available by the MIT-UH-IRTF Joint Campaign for NEO Reconnaissance. JH was supported by the grant 18-04514J of the Czech Science Foundation and by the Charles University Research program No. UNCE/SCI/023.
JH was supported by the grant 18-04514J of the Czech Science Foundation and by the Charles University Research program No. UNCE/SCI/023. We thank Nick Moskovitz, Ellen Howell, and an anonymous reviewer, among others, for useful conversations and correspondence, which greatly improved this manuscript. Any opinions, findings, and conclusions or recommendations expressed in this material are those of the author(s) and do not necessarily reflect the views of NASA.

Tables.

| Date (UTC) | Object Name | Airmass Range | Description | Wavelength Range | Phase Angle (degrees) | Rotational Coverage |
|---|---|---|---|---|---|---|
| 11/05/2007 | SAO 93936 | 1.12 | Visible Solar Analogue | 0.4-0.74 | | |
| "" 9:21-12:50 UTC | (3200) Phaethon | 1.22 - 1.00 | Asteroid | "" | 40.8 | 0.94 rotations |
| 12/12/2017 | SAO 39829 | 1.793 - 1.101 | NIR Standard Star | 0.68-2.55 | | |
| "" | SAO 93936 | 1.006 | NIR Solar Analogue | "" | | |
| "" 6:01-9:59 UTC | (3200) Phaethon | 1.576 - 1.109 | Asteroid | "" | 21.8 | 1.09 rotations |

Table 1. Log of observations of asteroid targets and calibration stars for our 2007 visible observations and our 2017 near-infrared observations.

Figures.

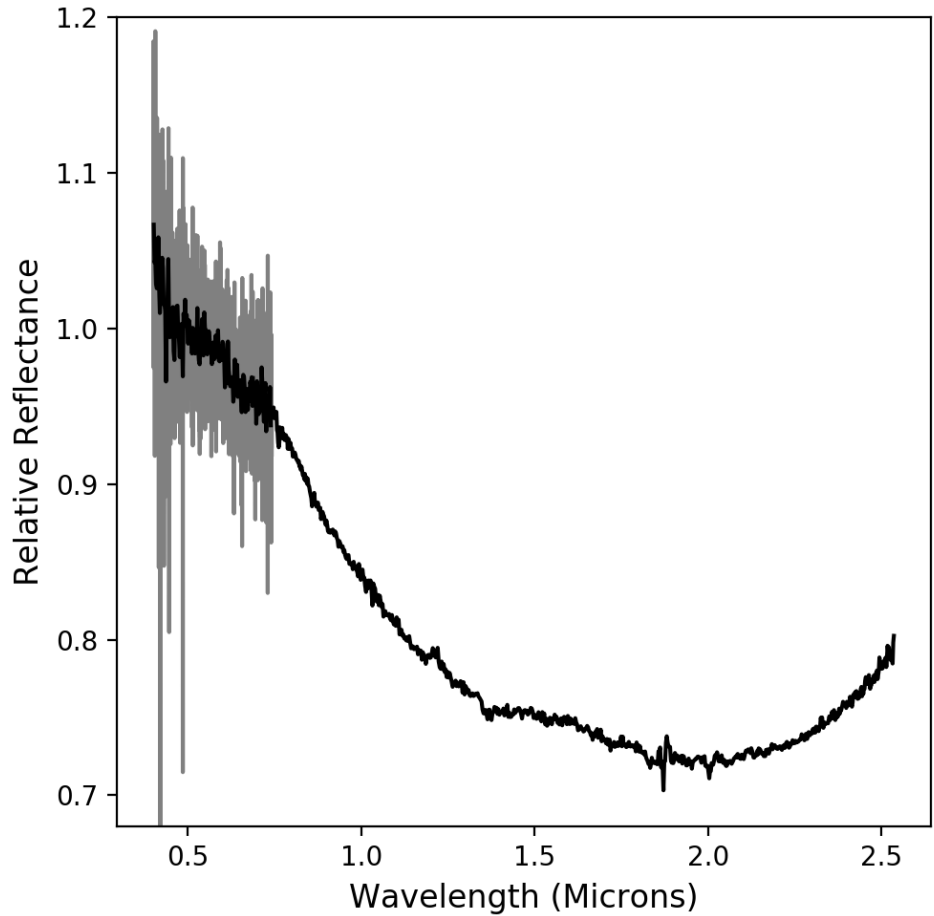

Figure 1. Rotationally-averaged Visible-NIR data (0.4-2.55 µm) of (3200) Phaethon showing a mostly blue and featureless spectrum. The upturn in reflectance values past 2.0 µm is due to the shorter wavelength end of the Planck curve that is shifted to NIR wavelengths due to the asteroids low heliocentric distance (higher surface temperature) at the time of the observations. The visible observations are binned to a comparable NIR spectrum (black) resolution, and the grey data are the unbinned visible data. The visible data are of higher resolution and were obtained when Phaethon was fainter (V. Mag. 16.37).

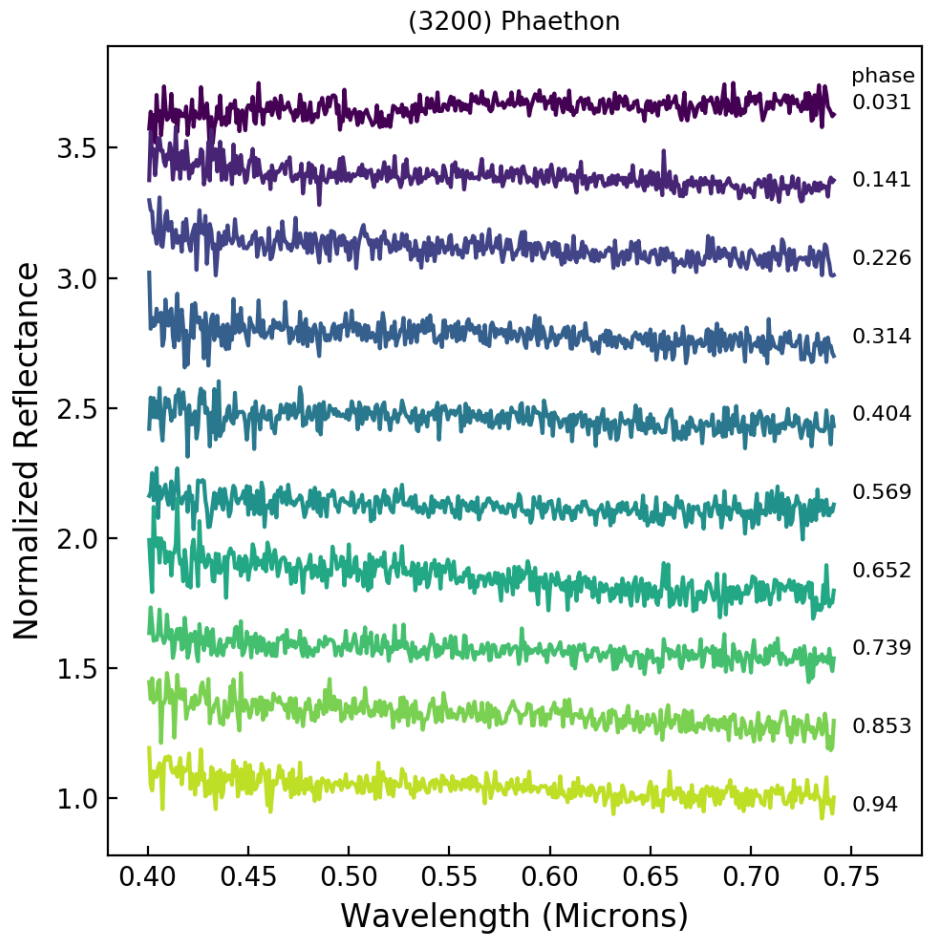

Figure 2. Visible wavelength spectra of Phaethon obtained at different rotational phases (shown to the right of each spectrum) showing the mostly blue spectral slope. The spectra are offset in reflectance for clarity. The weakly red spectrum near ~0.03 phase is likely different due to changing atmospheric conditions, and should be interpreted carefully.

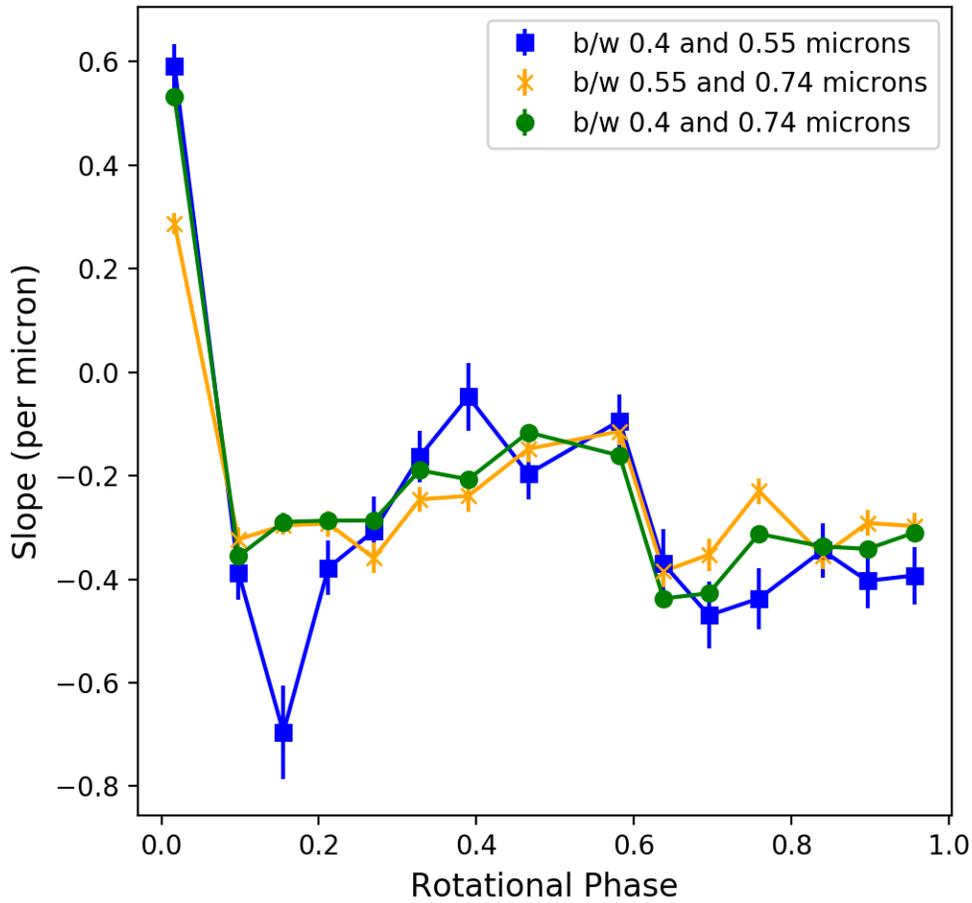

Figure 3. A plot showing the visible spectral slope defined between three possible wavelength ranges as a function of rotation phase. The green circles are a fit of all visible data, the blue squares are the data short of 0.55 μm, and the yellow Xs are for wavelengths longer than 0.55 μm. The vertical error bars show the 1-sigma uncertainty. The mean slope values are (0.4-0.55): -0.27 +/- 0.02, (0.55-0.74): -0.243+/-0.008, (0.4-0.74): -0.235 +/- 0.004. We see no evidence for a UV absorption feature – but we do see significant slope changes across the surface for all wavelength ranges considered.

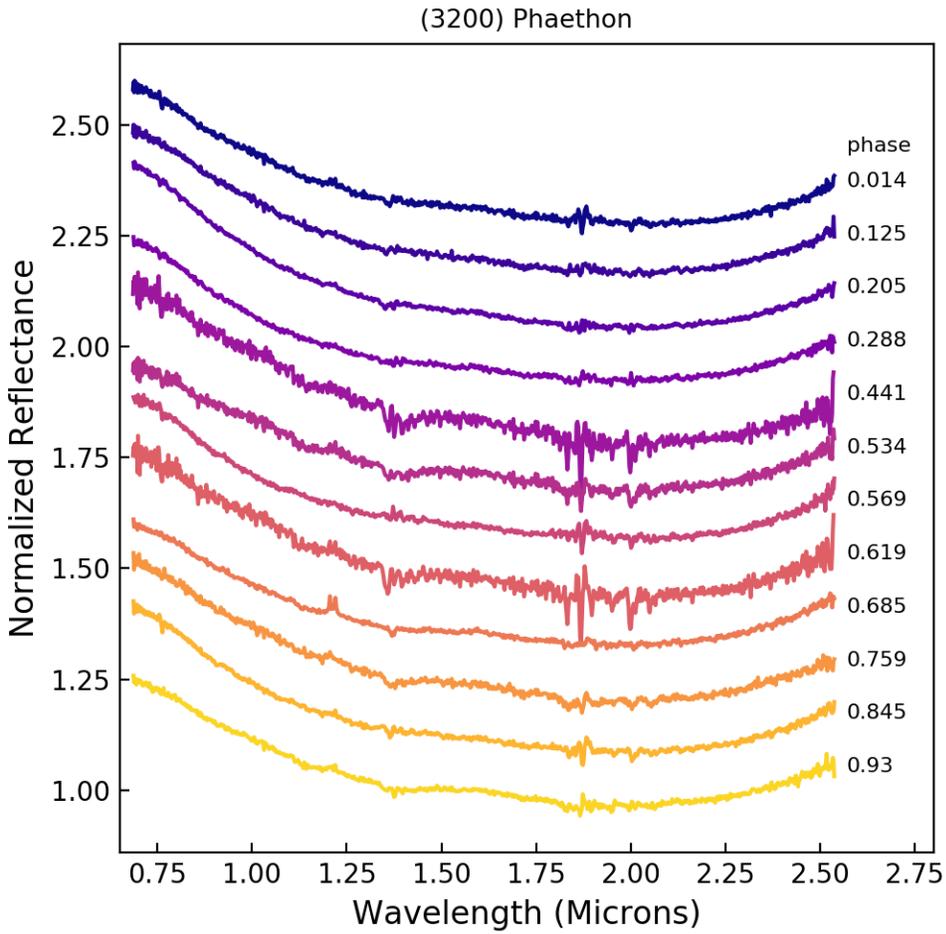

Figure 4. Near-IR spectra of Phaethon obtained using the NASA IRTF at different rotational phases (shown to the right of each spectrum) showing mostly blue spectral slope and a sharp rise in reflectance beyond 2.0 μm due to thermal emission. The spectra are offset in reflectance for clarity.

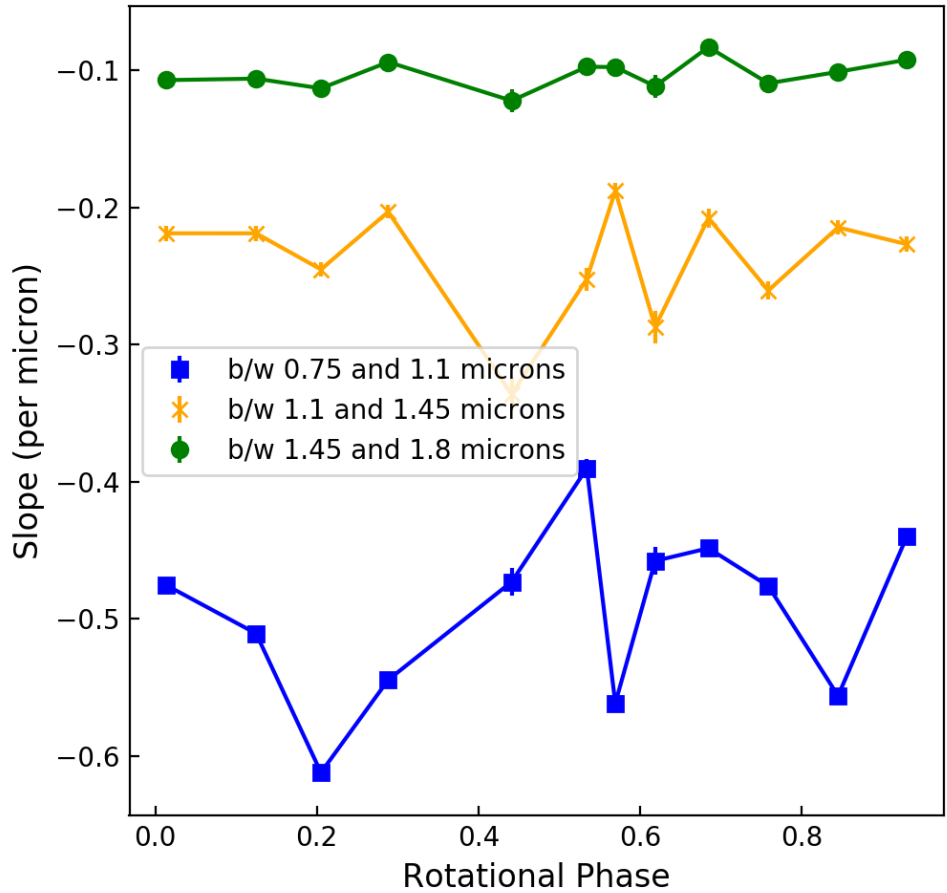

Figure 5. A plot showing the NIR spectral slope defined between three possible wavelength ranges as a function of rotation phase. The blue squares are the fit between 0.75 and 1.1 μm, the yellow Xs between 1.1 and 1.45 μm, and the green circles are the fits between 1.45 and 1.8 μm. The vertical error bars show the 1-sigma uncertainty. The curvature of the spectrum is clear in that the shorter wavelengths are much bluer than those at longer wavelengths. Additionally, the shorter wavelength slopes are observed to vary more significantly than the longer wavelength slopes. The average slopes for each wavelength segment are: (0.75, 1.1): -0.496 +/- 0.002, (1.1, 1.45): -0.238 +/- 0.002, (1.45, 1.8): -0.103 +/- 0.001. The start and end of our observations correspond to ~0.5-0.6 phase, such that the difference between those two data points should be interpreted at least partially as an effect of changing airmass and observing conditions. There are variations seen within the dataset, but no clear trends are observed.

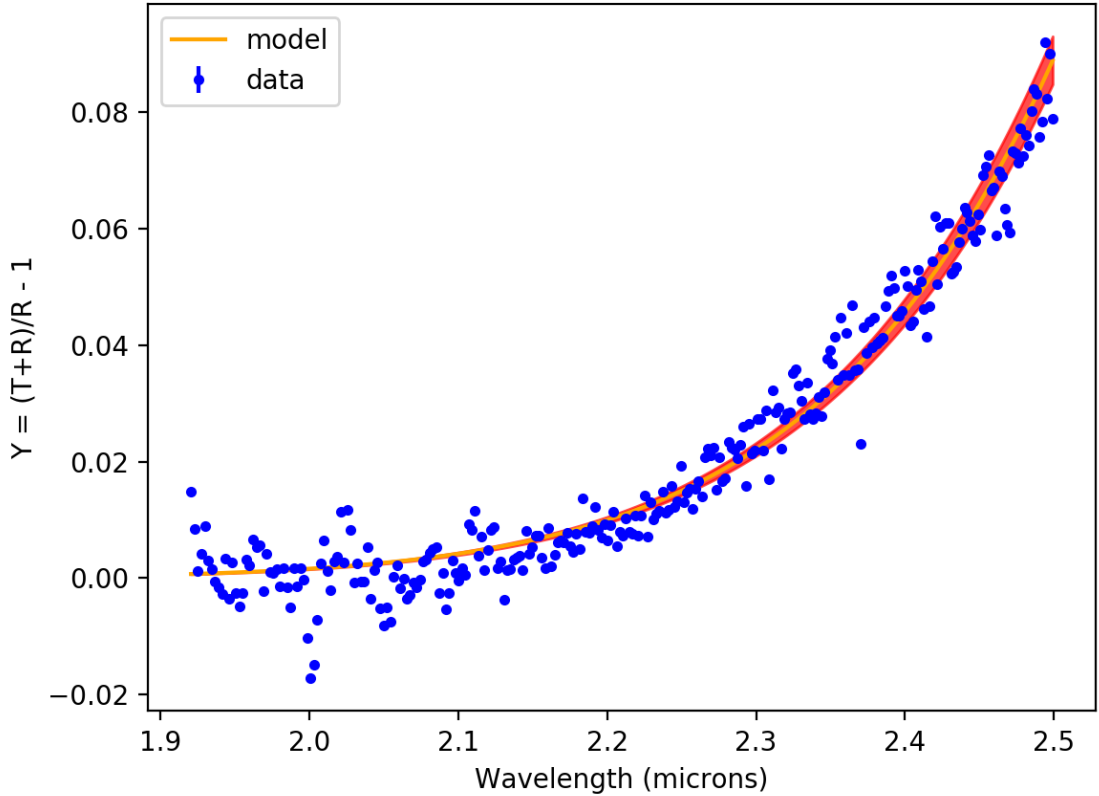

Figure 6. Plot shows a continuum removed the thermal tail of the NIR spectrum along with the model. The red shaded area represents the 1-sigma variation in albedo in the model. The modeled reflectance slope throughout the k-band is -0.015/μm, the ratio of k-band to v-band albedos is 0.716, and the best-fit albedo $p_v$ = 0.09 +/- 0.03 and beaming parameter $\eta$ = 1.65 +/- 0.14. This is slightly higher than our average values of $p_v$ = 0.08 +/- 0.01 and $\eta$ = 1.70 +/- 0.05. The reflectance continuum was chosen as the best-fit slope of the data between 1.95 and 2.1 microns.

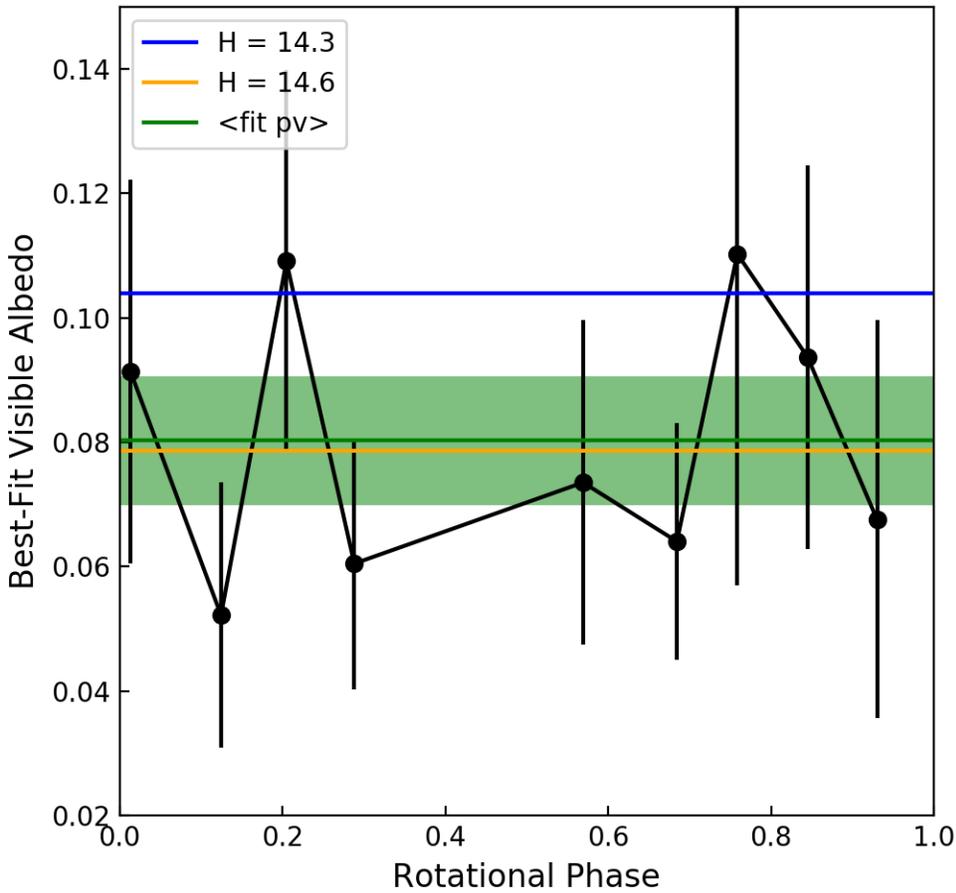

Figure 7. A plot showing the change in albedo as a function of rotation phase for Phaethon. The black circles are the albedos we estimated for Phaethon by modeling the thermal emission. The error bars show 1-sigma uncertainties. The blue and yellow horizontal bars correspond to the visible albedos one would expect for the two reported absolute magnitudes given the recently derived radar diameter (~5.7 km, Taylor et al. 2018). The green horizontal line is the average of our nine best-fit visible albedos (assuming G=0.06, Ansdell et al., 2014), $p_v$ = 0.08 +/- 0.01, with the green zone representing a 1-sigma variation from that mean. The individual albedos vary from 0.052 to 0.011. Almost all previous radiometric measurements of Phaethon's albedo (and diameter) would plot at the blue (H = 14.3) line or above.